\documentstyle[preprint,aps]{revtex}

\begin{document}

\draft


\tighten

\title{Vector Supersymmetry and Finite Quantum Correction of Chern-Simons
Theory in the Light-Cone Gauge}

\author{ W.F. Chen\renewcommand{\thefootnote}{\dagger}\footnote{
E-mail: wchen@msnet.mathstat.uoguelph.ca} and G. Leibbrandt
\renewcommand{\thefootnote}{\ddagger}\footnote{ E-mail:
gleibbra@msnet.mathstat.uoguelph.ca}}

\address{ Department of Mathematics and Statistics,
University of Guelph\\ Guelph, Ontario, N1G 2W1, Canada}

\maketitle

\begin{abstract}
 Vector supersymmetry is shown to exist also in light-cone gauge
 Chern-Simons theory.  Using a gauge
 invariant regularization scheme, we demonstrate explicitly
 that the finite quantum correction to
 the coupling constant of Chern-Simons theory
 is intimately associated with the breaking of vector
 supersymmetry at the regularization level.
 The advantage of investigating such a quantum phenomenon in
 the light-cone gauge is emphasized and
the BRST and vector supersymmetry invariance of quantum effective
action is discussed.

 \vspace{3ex}
{\noindent $PACS$: 11.10.Kk; 11.15.-q}\\
{\noindent $Keywords$: Chern-Simons theory; light-cone gauge;
vector supersymmetry; finite quantum correction.}
\end{abstract}

\vspace{3ex}

\begin{flushleft}
{\bf 1.~Introduction}
\end{flushleft}

\vspace{2mm}

Gauge fixing is an indispensable procedure in quantizing a gauge
field theory. Different gauge choices define distinct
hypersurfaces which intersect the gauge orbits in the
configuration space of the gauge field and lead to different
gauge-fixed effective actions. The gauge-fixed effective action
breaks the local gauge symmetry but preserves a rigid BRST
symmetry. The latter is known to play a fundamental role, since it
ensures not only the renormalizability and unitarity of the
underlying gauge theory, but also guarantees its gauge
independence. In addition to this rigid BRST symmetry, there
exists a BRST-like vector supersymmetry which was discovered in
three-dimensional Chern-Simons theory and which can arise with
certain gauge choices, such as the Landau gauge \cite{brt}.
Although this vector supersymmetry survives only in a particular
gauge, it was nevertheless beneficial in analyzing the dynamical
structure of Chern-Simons theory, for instance in proving its
perturbative finiteness \cite{delduc,dado}. Moreover, it was
argued that this vector supersymmetry may not only play a role in
constructing physical observables and, therefore, possesses actual
physical relevance \cite{gms}, but might also be the symmetry
origin for the infrared safety of topologically massive Yang-Mills
theory \cite{chen}. For these very reasons it would be desirable
to clarify the physical status of vector supersymmetry.

 The dynamical effects of vector supersymmetry
 in the Landau gauge have been investigated by various authors
 \cite{delduc,dado,gmr}.
 They showed that Landau
 vector supersymmetry imposes subtle constraints
 on quantum Chern-Simons gauge theory \cite{delduc,dado},
 preventing the theory from receiving any quantum corrections
 and thereby keeping the quantum theory identical to its classical
 counterpart. So much for the theoretical end
 of things. But what happens to vector supersymmetry in an actual
 calculation, i.e. in a perturbative calculation, which is known
 to require an intermediate regularization procedure in order to
 handle the UV divergence? The question is nontrivial since we
 know that BRST symmetry and Landau vector supersymmetry
 cannot survive simultaneously at the regularization level.
In short, there exists no regularization procedure preserving both
the BRST symmetry and Landau vector supersymmetry. Moreover, if a
regularization scheme preserves BRST symmetry, but breaks Landau
vector supersymmetry, the coupling constant of Chern-Simons theory
will receive a finite quantum correction \cite{col,do,ma1}. By
contrast, if the Landau vector supersymmetry is respected and the
BRST symmetry violated, there is no quantum correction and the
coupling constant keeps its classical value \cite{gmm,csw}.
Accordingly, since BRST symmetry is the most fundamental symmetry
a gauge theory can possess, it seems reasonable to work with a
regularization scheme that will preserve that symmetry \cite{gmr}.

A few years ago researchers discovered --- again in the framework
of Chern-Simons theory --- that vector supersymmetry manifests
itself in noncovariant gauges as well, specifically in gauges of
the axial type \cite{ma,brand}. A detailed practical calculation
was performed in the light-cone gauge by the authors of
Ref.\,\cite{lema}. Employing a BRST-invariant regularization,
consisting of higher covariant-derivative regularization and
dimensional regularization, these authors demonstrated the
appearance of a finite quantum correction to the coupling
constant. This result motivated us to check whether the finite
quantum correction is connected with the breaking of vector
supersymmetry at the regularization level.

\vspace{2mm}

\begin{flushleft}
{\bf 2. Symmetries in light-cone Chern-Simons theory}
\end{flushleft}

\vspace{2mm}

 The classical Euclidean action of $SU(N)$ Chern-Simons
theory reads
\begin{eqnarray}
S=-\frac{ik}{4\pi}\int
d^3x\,\epsilon_{\mu\rho\nu}\left(\frac{1}{2} A_\mu^a\partial_\rho
A_\nu^a+\frac{1}{3!}f^{abc}A_\mu^a A_\rho^bA_\nu^c\right),
~~\mu,\nu,\rho=1,2,3,
\end{eqnarray}
where $f^{abc}$ are the structure constants of the gauge group
$SU(N)$ and $k$ is the bare statistical parameter of Chern-Simons
theory. It is convenient to write $|k|=4\pi/g^2$ and to rescale
the gauge field $A^a_\mu$ as $A\rightarrow gA$. Hence, in the
homogeneous light-cone gauge $n\cdot A^a=0$, $n^2=0$, the
gauge-fixed Chern-Simons action becomes
\begin{eqnarray}
S&=&-i\,\mbox{sign}(k)\int
d^3x\,\epsilon_{\mu\rho\nu}\left(\frac{1}{2} A_\mu^a\partial_\rho
A_\nu^a+\frac{1}{3!}g f^{abc}A_\mu^a
A_\rho^bA_\nu^c\right)\nonumber\\
&&+\int d^3x\left(B^a n_\mu A_\mu^a+\bar{c}^a n_\mu
D_{\mu}^{ab}c^b\right), \label{action}
\end{eqnarray}
where $D_\mu^{ab}=\partial_\mu\delta^{ab}+gf^{abc}A_\mu^c$ is the
covariant derivative.  As in four-dimensional Yang-Mills, the
theory possesses the usual BRST symmetry, namely,
\begin{eqnarray}
sA_\mu^a=D_\mu^{ab}c^b,~~sc^a=-\frac{1}{2}gf^{abc}c^bc^c,~~
s\bar{c}^a=B^a, ~~sB^a=0. \label{brst}
\end{eqnarray}
In addition,  Chern-Simons theory exhibits, just as in the Landau
gauge, the vector supersymmetry
\begin{eqnarray}
v_\nu A_\mu^a=i\,\mbox{sign}(k)\epsilon_{\nu\mu\alpha}n_\alpha
c^a, ~~v_\nu c^a=0, ~~v_\nu \overline{c}^a=A_\nu^a,~~v_\nu
B^a=-\partial_\nu c^a, \label{vector1}
\end{eqnarray}
and the anti-vector supersymmetry
\begin{eqnarray}
\overline{v}_\nu A_\mu^a=i\,\mbox{sign}(k)
\epsilon_{\nu\mu\alpha}n_\alpha \overline{c}^a, ~~
\overline{v}_\nu \overline{c}^a=0, ~~\overline{v}_\nu
c^a=-A_\nu^a,~~\overline{v}_\nu B^a=-\partial_\nu \overline{c}^a,
\label{vector2}
\end{eqnarray}
leading to the following supersymmetric variations,
\begin{eqnarray}
v_\nu S&=&\int d^3x\left[\partial_\rho\left(n_\rho c^a
A_\nu^a\right)-\partial_\nu\left(n_\rho c^a
A_\rho^a\right)\right]=0;\nonumber\\
\overline{v}_\nu S&=&\int d^3x\left[\partial_\rho\left(n_\rho
\overline{c}^a A_\nu^a\right)-\partial_\nu\left(n_\rho
\overline{c}^a A_\rho^a\right)\right]=0. \label{inv}
\end{eqnarray}

\vspace{2mm}

\begin{flushleft}
{\bf 3. Breaking of vector supersymmetry in BRST-invariant
regularization}
\end{flushleft}

\vspace{2mm}

Let us first analyze how the supersymmetries in
Eqs.\,(\ref{vector1}) and (\ref{vector2}) are broken in a
BRST-invariant regularization method. As we know, dimensional
regularization is generally regarded as a reliable and convenient
regularization scheme, preserving gauge invariance in a natural
way. Of course, it is also well known that the presence of
$\gamma_5$ or, equivalently, of the epsilon tensor, requires a
special treatment, such as  the consistent dimensional
regularization scheme proposed by 't Hooft and Veltman
\cite{chai,hooft}. Unfortunately, as shown in Ref.\,\cite{ma1}, a
consistent dimensional regularization continuation does not exist
for pure Chern-Simons gauge theory, since the three-dimensional
antisymmetric tensor $\epsilon_{\mu\nu\rho}$ prevents inversion of
the kinetic term in $d$-dimensional space, even after
gauge-fixing. To obtain a regularized Chern-Simons theory,
consistent with gauge symmetry, one must adopt a kind of hybrid
regularization ---- a combination of higher covariant-derivative
regularization and consistent dimensional regularization. The
simplest higher covariant expression is the three-dimensional
Yang-Mills term $S_{\rm YM}$,
\begin{eqnarray}
S_{\rm YM}=\frac{1}{4m}\int d^3x\, F_{\mu\nu}^aF^a_{\mu\nu},
\label{ym}
\end{eqnarray}
$m$ being the regulator mass. The $S_{\rm YM}$ term should be
added to the Chern-Simons action {\it before} performing the
dimensional continuation according to the 't Hooft-Veltman
prescription \cite{ma1,lema,chai,hooft}. The regularized theory
has now two regulators: the dimensional parameter $\epsilon=3-d$
and the mass regulator $m$, the order of taking the limits being
first $d\rightarrow 3$, then $m\rightarrow \infty$.

Although  the BRST symmetry (\ref{brst}) is now preserved at the
regularization level, the introduction of the Yang-Mills term
$S_{\rm YM}$ violates the supersymmetries (\ref{vector1}) and
(\ref{vector2}) explicitly. A straightforward calculation gives
\begin{eqnarray}
v_\nu S_{\rm YM}&=&v_\nu^{(0)}+v_\nu^{(1)}+v_\nu^{(2)},
\nonumber\\
v_\nu^{(0)}&=& \frac{i}{m}\int d^3x\, \mbox{sign} (k)
\epsilon_{\nu\alpha\beta}n_\beta\, \partial_\mu
c^a\left(\partial_\mu
A_\alpha^a-\partial_\alpha A_\mu^a\right),\nonumber\\
v_\nu^{(1)}&=& \frac{i}{m}\int d^3x\,\mbox{sign} (k)
\epsilon_{\nu\alpha\beta}n_\beta\, gf^{abc}\left[\partial_\mu c^a
A_\mu^b A_\alpha^c +\left(\partial_\mu A_\alpha^a-\partial_\alpha
A_\mu^a\right)A_\mu^bc^c\right],\nonumber\\
 v_\nu^{(2)}&=& \frac{i}{m}\int d^3x\,\mbox{sign} (k)
\epsilon_{\nu\alpha\beta}n_\beta\, g^2f^{abc}f^{ade}
A_\mu^bc^cA_\mu^dA_\alpha^e\, , \label{ymv1}
\end{eqnarray}
and
\begin{eqnarray}
\overline{v}_\nu S_{\rm YM}&=& \overline{v}_\nu^{(0)}+ \overline{
v}_\nu^{(1)}+ \overline{v}_\nu^{(2)},
\nonumber\\
\overline{v}_\nu^{(0)}&=& \frac{i}{m}\int d^3x \,\mbox{sign} (k)
\epsilon_{\nu\alpha\beta}n_\beta\, \partial_\mu
\overline{c}^a\left(\partial_\mu
A_\alpha^a-\partial_\alpha A_\mu^a\right),\nonumber\\
\overline{v}_\nu^{(1)}&=& \frac{i}{m}\int d^3x\, \mbox{sign} (k)
\epsilon_{\nu\alpha\beta}n_\beta\, gf^{abc}\left[\partial_\mu
\overline{c}^a A_\mu^b A_\alpha^c +\left(\partial_\mu
A_\alpha^a-\partial_\alpha
A_\mu^a\right)A_\mu^b\overline{c}^c\right],\nonumber\\
\overline{v}_\nu^{(2)}&=& \frac{i}{m}\int d^3x\, \mbox{sign} (k)
\epsilon_{\nu\alpha\beta}n_\beta\, g^2f^{abc}f^{ade} A_\mu^b
\overline{c}^cA_\mu^dA_\alpha^e\, . \label{ymv2}
\end{eqnarray}

 To find out whether the finite quantum correction comes from
 the breaking of vector supersymmetry at the regularization level,
 we start from the corresponding broken Ward
 identity and derive
 relations among the various Green functions.
We shall concentrate on the supersymmetry in Eq.\,(\ref{vector1}).
The standard way of obtaining the Ward identity is to
 introduce external sources for the fields and their
 symmetric transformations.
Occasionally, however, as in the present case, it is possible to
extract the required Ward identity in a shorter, more efficient
way, by considering a general functional $F[\Phi]$ of the field
$\Phi=\left(A_\mu^a,B^a,c^a,\bar{c}^a\right)$. Invariance of the
quantum observable
\begin{eqnarray}
\langle F[\Phi]\rangle =\int {\cal D}{\Phi}F[\Phi]e^{-S[\Phi]}
\label{obv}
\end{eqnarray}
under an arbitrary infinitesimal transformation
$\Phi\rightarrow\Phi +\delta\Phi$ then yields the following
identity,
\begin{eqnarray}
\left\langle \frac{\partial F[\Phi]}{\partial \Phi}\delta\Phi
-\delta S[\Phi]F[\Phi]\right\rangle =0. \label{identity}
\end{eqnarray}
 Choosing $F[\Phi]$ to be of the form $F[\Phi]=A_\mu^a (x)\bar{c}^b(y)$,
one can prove from Eq.\,(\ref{identity}) that the vector
supersymmetry transformation (\ref{vector1}) leads to the vacuum
expectation value,
\begin{eqnarray}
\left\langle A_\mu^a (x)A_\nu^b (y)\right\rangle =i \,\mbox{sign}
(k) \epsilon_{\mu\nu\rho}n_\rho\left\langle
c^a(x)\overline{c}^b(y)\right\rangle +\left\langle A_\mu^a
(x)\overline{c}^b(y) v_\nu S_{\rm YM}\right\rangle . \label{green}
\end{eqnarray}
The broken vector supersymmetry manifests itself through the
second term on the R.H.S. of Eq.\,(\ref{green}), which is seen to
contain both the gauge field $A^a_\mu$ and the anti-ghost field
$\overline{c}^b$. In momentum space, Eq.\,(\ref{green}) reads
\begin{eqnarray}
G_{\mu\nu}(p)=i\,\mbox{sign} (k)\epsilon_{\mu\nu\rho} n_{\rho}
S(p) +G_{\mu\rho}(p)\Omega_{\rho\nu}(p)S(p), \label{relation}
\end{eqnarray}
where $G_{\mu\nu}(p)$ and $S(p)$ denote the propagators for the
gauge field and ghost field in momentum space, respectively;
$\Omega_{\mu\nu}(p)$ is related to the composite operator $v_\nu
S_{\rm YM}$ and is the 1PI part of the Green function
$\left\langle A_\mu^a (x)\overline{c}^b(y) v_\nu S_{\rm
YM}\right\rangle$ in momentum space, namely,
\begin{eqnarray}
\left\langle A_\mu^a (x)\bar{c}^b(y) v_\nu S_{\rm YM}\right\rangle
&=& \int d^3u d^3v G_{\mu\rho}(x-u)\Omega_{\rho\nu} (u-v) S(v-y)
\nonumber\\
&=&\int\frac{d^3p}{(2\pi)^3}G_{\mu\rho}(p)\Omega_{\rho\nu}(p)S(p)
e^{ip\cdot (x-y)}.
\end{eqnarray}

\vspace{2mm}

\begin{flushleft}
{\bf 4. Finite quantum correction as breaking effect of vector
supersymmetry}
\end{flushleft}

\vspace{2mm}

 In the following, we shall check identity (\ref{relation})
up to the one-loop level in order to convince ourselves that the
finite quantum correction  is indeed associated with the breaking
of vector supersymmetry. To this effect,  we first expand the
various quantities in (\ref{relation}) to one-loop order,
\begin{eqnarray}
G_{\mu\nu}(p)&=&G^{(0)}_{\mu\nu}(p)+G^{(0)}_{\mu\lambda}(p)
\Pi_{\lambda\rho}(p)G^{(0)}_{\rho\nu}(p)+{\cal O}(g^4); \nonumber\\
S(p)&=&S^{(0)}(p)+S^{(0)}(p)\Sigma (p)S^{(0)}(p)+{\cal O}(g^4);\nonumber\\
\Omega_{\mu\nu}(p)&=&\Omega^{(0)}_{\mu\nu}(p)+\Omega^{(1)}_{\mu\nu}(p)
+{\cal O}(g^4), \label{expansion}
\end{eqnarray}
where $G^{(0)}_{\mu\nu}(p)$ and $S^{(0)}(p)$ are the tree-level
propagators of the Chern-Simons gauge field and ghost field,
respectively. In Euclidean space-time \cite{lema},
$$\addtocounter{equation}{1}
G^{(0)}_{\mu\nu}(p)=\frac{m}{p^2(p^2+m^2)}\left[-m\,
\mbox{sign}(k)\epsilon_{\mu\nu\rho}p_\rho+\frac{m}{n\cdot p}
\mbox{sign}(k)\left(p_\mu
\epsilon_{\nu\lambda\rho}p_{\lambda}n_\rho
-p_\nu\epsilon_{\mu\lambda\rho}p_\lambda n_\rho\right)\right. $$
$$\quad{}+\left. p^2\delta_{\mu\nu} -\frac{p^2}{n\cdot p}\left(p_\mu
n_\nu +p_\nu n_\mu\right)\right], $$
 $$\quad{}=\frac{m}{(p^2+m^2)(n\cdot
 p)}\left[-m\,\mbox{sign}(k)\epsilon_{\mu\nu\rho}n_\rho
 +n\cdot p\delta_{\mu\nu}
 -\left(p_\mu n_\nu +p_\nu n_\mu\right)\right],
 \eqno{(16a)} \label{progluon} $$
$$S^{(0)}(p)=\frac{i}{n\cdot p}. \eqno{(16b)} \label{proghost}
$$
Here, we have employed the identity \cite{lema}
\begin{eqnarray}
\frac{1}{n\cdot p}\epsilon_{\mu\nu\rho}n_\rho =\frac{1}{p^2}
\left[\epsilon_{\mu\nu\rho}p_\rho -\frac{1}{n\cdot p} \left(
 p_\mu p_\rho n_\lambda\epsilon_{\nu\rho\lambda}-
 p_\nu p_\rho n_\lambda\epsilon_{\mu\rho\lambda}\right)\right].
\label{identity1}
\end{eqnarray}

The above propagators are seen to contain the spurious light-cone
gauge singularity $\left(n\cdot p\right)^{-1}$, which can be
treated by the following prescription \cite{mand,leib1,leib2} in
Minkowski space:
\begin{eqnarray}
\frac{1}{p\cdot n}&=&\lim_{\epsilon\rightarrow 0}\frac{p\cdot
n^*}{\left(p\cdot n\right)\left(p\cdot n^*\right)+i\epsilon},
~~~\epsilon >0,\nonumber\\
n_\mu &=&(n_0, {\bf n}), ~~n_\mu^*=(n_0,-{\bf n}), ~~n^2=n^{*2}=0.
  \label{presc}
\end{eqnarray}

In order to verify identity (\ref{relation}), we require not only
the one-loop vacuum polarization tensor $\Pi_{\mu\nu}(p)$ and the
ghost field self-energy $\Sigma (p)$, but also the 3-point and
4-point vertices of topologically massive Yang-Mills theory
(Figs.\,1a and 1b), namely,
$$\addtocounter{equation}{1}
V_{\mu\nu\rho}^{abc}(p,q,r)=igf^{abc}\left\{\mbox{sign}(k)
\epsilon_{\mu\nu\rho}+\frac{1}{m}\left[(p-q)_{\rho}\delta_{\mu\nu}
+(q-r)_{\mu}\delta_{\nu\rho} +(r-p)_{\nu}\delta_{\rho\mu} \right]
\right\} $$
$$\quad{}{\equiv}ig f^{abc}\left[\mbox{sign}(k) \epsilon_{\mu\nu\rho}+
\frac{1}{m}\widetilde{V}_{\mu\nu\rho}(p,q,r)\right] ,
~~p+q+r=0;\eqno{(19a)}\label{vertex1} $$
$$V_{\mu\nu\lambda\rho}^{abcd}(p,q,r,s)=-\frac{g^2}{m}\left[
f^{eab}f^{ecd}\left(\delta_{\mu\lambda}\delta_{\nu\rho}-
\delta_{\nu\lambda}\delta_{\rho\mu}\right) +
f^{eac}f^{ebd}\left(\delta_{\lambda\rho}\delta_{\mu\nu}-
\delta_{\nu\lambda}\delta_{\rho\mu}\right)\right. $$
$$\quad{}+\left. f^{ebc}f^{eda}\left(\delta_{\mu\lambda}\delta_{\nu\rho}-
\delta_{\mu\nu}\delta_{\lambda\rho}\right)\right],~~~ p+q+r+s=0,
\eqno{(19b)}\label{vertex2} $$ along with the ghost-ghost-gauge
vertex (Fig.\,1c),
$$
V_\mu^{abc}=-gf^{abc}n_\mu. \eqno{(19c)} \label{vertex3} $$
$\Omega_{\mu\nu}^{(0)}(p)$ and $\Omega_{\mu\nu}^{(1)}(p)$ may be
evaluated with the help of the Feynman rules for the composite
operators (cf. Eq.\,(\ref{ymv1}), see Fig.\,2):
\begin{eqnarray}
v^{(0)}_\nu: &&
\widetilde{V}^{ab}_{\nu\mu}(p,q)=\frac{i}{m}\mbox{sign}(k)
\epsilon_{\nu\alpha\beta}
n_{\beta}\delta^{ab}\left(p^2\delta_{\alpha\mu}-p_\alpha
p_\mu\right),
~~p+q=0; \\
v^{(1)}_\nu: && \widetilde{V}^{abc}_{\nu\mu\rho}(r,p,q)
=-\frac{1}{m}\mbox{sign}(k) \epsilon_{\nu\alpha\beta}n_\beta
gf^{abc}\left[\delta_{\mu\rho}
\left(p_\alpha-q_\alpha\right)+\delta_{\rho\alpha}
\left(q_\mu-r_\mu\right)\right.\nonumber\\
&& \left.+\delta_{\alpha\mu}
\left(r_\rho-p_\rho\right)\right]=-\frac{1}{m}
\mbox{sign}(k)\epsilon_{\nu\alpha\beta}n_\beta gf^{abc}
\widetilde{V}_{\alpha\mu\rho}(r,p,q),~~~p+q+r=0;\\
v^{(2)}_\nu: && \widetilde{V}^{dabc}_{\nu\mu\rho\lambda}(s,p,q,r)
=-\frac{i}{m}\mbox{sign}(k) \epsilon_{\nu\alpha\beta}n_\beta
g^2\left[
f^{eda}f^{ebc}\left(\delta_{\mu\rho}\delta_{\lambda\alpha}-
\delta_{\mu\lambda}\delta_{\rho\alpha}\right) \right.\nonumber\\
&&+\left.
f^{eac}f^{edb}\left(\delta_{\mu\rho}\delta_{\lambda\alpha}-
\delta_{\lambda\rho}\delta_{\mu\alpha}\right)+
f^{edc}f^{eab}\left(\delta_{\lambda\mu}\delta_{\rho\alpha}-
\delta_{\mu\alpha}\delta_{\lambda\rho}\right)\right] \nonumber\\
&&=i\,\mbox{sign}(k)\epsilon_{\nu\alpha\beta}n_\beta
V^{dabc}_{\alpha\mu\rho\lambda} (s,p,q,r),~~~ p+q+r+s=0.
\label{co}
\end{eqnarray}
Inserting the expansions (\ref{expansion}) into the identity
(\ref{relation}), we obtain both the tree-level relation
\begin{eqnarray}
G^{(0)}_{\mu\nu}(p)=i\,\mbox{sign}(k)\epsilon_{\mu\nu\rho}n_\rho
S^{(0)}(p) +
G^{(0)}_{\mu\rho}(p)\Omega_{\rho\nu}^{(0)}(p)S^{0}(p), \label{tre}
\end{eqnarray}
and the one-loop relation,
\begin{eqnarray}
G^{(0)}_{\mu\lambda}(p)\Pi_{\lambda\rho}(p)G^{(0)}_{\rho\nu}(p)
&=&i\,\mbox{sign}(k)\epsilon_{\mu\nu\rho}n_\rho S^{(0)}(p)\Sigma
(p)S^{(0)}(p)\nonumber\\
&&+G^{(0)}_{\mu\rho}(p)\Omega_{\rho\nu}^{(0)}(p)
S^{(0)}(p)\Sigma(p)S^{(0)}(p)\nonumber\\
&&+G^{(0)}_{\mu\rho}(p)\Omega_{\rho\nu}^{(1)}(p) S^{(0)}(p)
\nonumber\\
&&+G^{(0)}_{\mu\sigma}(p)\Pi_{\sigma\lambda}(p)
G^{(0)}_{\lambda\rho}(p) \Omega_{\rho\nu}^{(0)}(p) S^{(0)}(p).
\label{lre}
\end{eqnarray}
The tree-level relation (\ref{tre}) is satisfied trivially, as can
be seen by applying the Feynman rules in Eqs.\,(16) and (19).

Our real interest lies with the one-loop relation (\ref{lre}) to
see whether or not the shift in the Chern-Simons parameter  $k$
does indeed arise from  the breaking of the vector supersymmetry
(\ref{vector1}). In order to  verify (\ref{lre}), we shall proceed
as follows. We shall begin  by calculating the one-loop quantities
$\Sigma (p)$ and $\Omega^{(1)}_{\mu\nu}(p)$, comparing our result
with the value of the vacuum polarization tensor $\Pi_{\mu\nu}$
obtained in Ref.\,\cite{lema}, and then ascertain whether or not
relation (\ref{lre}) is consistent with the aforementioned finite
quantum correction.

We now turn to the evaluation of $\Sigma (p)$ and
$\Omega^{(1)}_{\mu\nu}(p)$. Although the dimensionally continued
gauge propagator is fairly complicated due to the presence of the
$\epsilon_{\mu\nu\rho}$ tensor, a detailed analysis \cite{lema}
shows that the gauge propagator may actually be `` simplified " by
decomposing it into a $d$-dimensional part and an evanescent part.
The advantage of this separation  is that  the evanescent  portion
exhibits an improved $UV$ behaviour and consequently vanishes in
the limit as $d\rightarrow 3$. Accordingly, it is perfectly safe
to work with the propagator (16a). We also recall that in the
light-cone gauge,
\begin{eqnarray}
 n_\mu G^{(0)}_{\mu\nu}(p)=0,
 \label{transverse}
\end{eqnarray}
so that the ghost field self-energy (Fig.\,3), given by
\begin{eqnarray}
\Sigma (p)\delta^{ab}=\int \frac{d^dk}{(2\pi)^d}\left(-g
f^{acd}n_\mu\right)G^{(0)}_{\mu\nu}(k)\left(-g f^{bdc}n_\nu\right)
S^{(0)}(k+p),
\end{eqnarray}
actually vanishes:
\begin{eqnarray}
\Sigma (p)=0.
\end{eqnarray}
Moreover, the last term on the R.H.S. of Eq.\,(\ref{lre}) also
vanishes in the limit $m\rightarrow\infty$, since
$\Omega_{\mu\nu}^{(0)}(p)$ is proportional to $1/m$, while
$\Pi_{\mu\nu}(p)$ is finite in the large $m$ limit. Hence, the
one-loop identity (\ref{lre}) reduces to
\begin{eqnarray}
G^{(0)}_{\mu\lambda}(p)\Pi_{\lambda\rho}(p)G^{(0)}_{\rho\nu}(p)
=G^{(0)}_{\mu\rho}(p)\Omega_{\rho\nu}^{(1)} (p) S^{(0)}(p).
\label{identity2}
\end{eqnarray}

It remains to evaluate $\Omega_{\mu\nu}^{(1)} (p)$ and see whether
its value is consistent with the vacuum polarization obtained in
Ref.\,\cite{lema}, namely,
\begin{eqnarray}
\Pi_{\mu\nu}(p)&=&\frac{1}{8\pi}\mbox{sign}(k)c_V
g^2\left[4\epsilon_{\mu\nu\rho}p_\rho -\frac{3p\cdot n^*}{n\cdot
n^*}\epsilon_{\mu\nu\rho}n_\rho\right.\nonumber\\
&& \left.+\frac{3p\cdot n^*}{(n\cdot n^*)(p\cdot
n)}\left(n_\mu\epsilon_{\nu\lambda\rho}
-n_\nu\epsilon_{\mu\lambda\rho}\right)n_\lambda p_\rho\right].
\label{vapo}
\end{eqnarray}
The first term in Eq.\,(\ref{vapo}) contributes to the local
quantum effective action, while the third term proportional to
$p\cdot n^*$ gives a nonlocal contribution which may be absorbed
into the local effective action by a finite, non-multiplicative
wave function renormalization \cite{lema}.

As shown in Fig.\,4, $\Omega^{(1)}_{\mu\nu}(p)$ gives rise to five
one-loop diagrams, only two of which are non-zero (Figs.\,4a and
4b). The contribution from Fig.\,4a reads
\begin{eqnarray}
\Omega_{ {\rm (a)}\mu\nu}^{(1)ab}(p)
&=&\frac{1}{2}\left.\int\frac{d^dk}{(2\pi)^d}
\widetilde{V}^{bcda}_{\nu\rho\lambda\mu}(-p,k,-k,p)
G^{(0)}_{\rho\lambda}(k)\delta^{cd}\right|_{d\rightarrow 3}
\nonumber\\
&=&-\frac{i}{2}\mbox{sign}(k)g^2 c_V\delta^{ab}
\int\frac{d^dk}{(2\pi)^d}\frac{1}{n\cdot k (k^2+m^2)}\left[
\left(\epsilon_{\nu\rho\beta}\delta_{\mu\lambda}n_\beta +
\epsilon_{\nu\lambda\beta}\delta_{\rho\mu}n_\beta
\right.\right.\nonumber\\
&&\left.\left.\left.-2\delta_{\lambda\rho} \epsilon_{\nu\mu\beta}
n_{\beta}\right)\left(-m \mbox{sign}(k)
\epsilon_{\rho\lambda\alpha}n_\alpha+n\cdot
k\delta_{\rho\lambda}-k_\lambda n_\rho-k_\rho n_\lambda\right)
\right]\right|_{d\rightarrow 3}
\nonumber\\
&=&\left.i\,\mbox{sign}(k)g^2
c_V\delta^{ab}\epsilon_{\nu\lambda\beta}n_{\beta} n_\mu
\int\frac{d^dk}{(2\pi)^d}\frac{k_\lambda}{n\cdot k (k^2+m^2)}
\right|_{d\rightarrow 3}
\nonumber\\
 &=&\frac{i}{4\pi}g^2
c_V\delta^{ab}m \mbox{sign}(k)\epsilon_{\nu\lambda\beta}n_{\beta}
\frac{n_\mu n_\lambda^*}{n\cdot n^*}, \label{valuea}
\end{eqnarray}
where we have used prescription (\ref{presc}), together with the
formula (see Appendix of Ref.\cite{lema}),
\begin{eqnarray}
\lim_{d\rightarrow
3}\int\frac{d^dk}{(2\pi)^d}\frac{k_\lambda}{n\cdot k
(k^2+m^2)}=\frac{1}{4\pi}\,\frac{m n_\lambda^*}{n\cdot n^*};
\end{eqnarray}
$c_V$ is the quadratic Casimir operator in the adjoint
representation given by $f^{acd}f^{bcd}=c_V\delta^{ab}$.
Evaluation of $\Omega_{ {\rm (b)}\mu\nu}^{(1)ab}(p)$ in Fig. 4b is
somewhat lengthier. We find that
\begin{eqnarray}
&&\Omega_{{\rm (b)}\mu\nu}^{(1)ab}(p)
=\frac{1}{2}\int\frac{d^dk}{(2\pi)^d} igf^{adc} V_{\mu\lambda\rho}
(p,-k-p,k) G^{(0)}_{\lambda\alpha}(k)
\widetilde{V}_{\nu\alpha\sigma}^{bdc} (-p,-k,k+p)
G^{(0)}_{\sigma\rho}(k+p)
\nonumber\\
&&=\frac{i}{2}\mbox{sign}(k)g^2c_V\delta^{ab}
\int\frac{d^dk}{(2\pi)^d}V_{\mu\lambda\rho} (p,-k-p,k)
G^{(0)}_{\lambda\alpha}(k)G^{(0)}_{\sigma\rho}(k+p)
\left(-\frac{1}{m}\right)\epsilon_{\nu\beta\gamma}n_{\gamma}
\widetilde{V}_{\beta\alpha\sigma}\nonumber\\
&&=\frac{i}{2}\mbox{sign}(k)g^2c_V\delta^{ab}\int\frac{d^dk}{(2\pi)^d}
V_{\mu\lambda\rho}
(p,-k-p,k)G^{(0)}_{\lambda\alpha}(k)G^{(0)}_{\sigma\rho}(k+p)
\epsilon_{\nu\beta\gamma}n_{\gamma} \left[-V_{\beta\alpha\sigma}
+\mbox{sign}(k)\epsilon_{\beta\alpha\sigma}\right]\nonumber\\
&&=ig^2c_V\delta^{ab}\left\{-\mbox{sign}(k)
\epsilon_{\nu\beta\gamma} n_\gamma \Pi_{{\rm (b)}\mu\beta}(p)
+\frac{1}{2}\int\frac{d^dk}{(2\pi)^d}\left[ V_{\mu\lambda\rho}
(p,-k-p,k)G^{(0)}_{\lambda\alpha}(k)\right.\right.
\nonumber\\
&&\times \left.\left.
G^{(0)}_{\sigma\rho}(k+p)\epsilon_{\nu\beta\gamma} n_\gamma
\epsilon_{\beta\alpha\sigma}\right]\frac{}{}\right\}. \label{ome1}
\end{eqnarray}
The second term in Eq.\,(\ref{ome1}) vanishes identically, since
$n_\mu G_{\mu\nu}=0$, i.e.
\begin{eqnarray}
&& \int\frac{d^dk}{(2\pi)^d}V_{\mu\lambda\rho}
(p,-k-p,k)G^{(0)}_{\lambda\alpha}(k)G^{(0)}_{\sigma\rho}(k+p)
\epsilon_{\nu\beta\gamma} n_\gamma\epsilon_{\beta\alpha\sigma}
\nonumber\\
&=&\int\frac{d^dk}{(2\pi)^d}V_{\mu\lambda\rho}
(p,-k-p,k)G^{(0)}_{\lambda\alpha}(k)G^{(0)}_{\sigma\rho}(k+p)
\left(\delta_{\nu\sigma}n_\alpha-\delta_{\nu\alpha}n_\sigma\right)
=0.
\end{eqnarray}
The next step is to insert the value for $\Pi_{{\rm
(b)}\mu\beta}(p)$, which represents the vacuum polarization tensor
from a gluon loop with two three-gauge vertices and two gauge
propagators. In the limits as $d\rightarrow 3$ and
$m\rightarrow\infty$,
 $\Pi_{{\rm (b)}\mu\beta}(p)$ becomes \cite{lema}
\begin{eqnarray}
\Pi_{{\rm (b)}\mu\beta}(p)&=& \frac{i}{8\pi} c_Vg^2\left[
4\,\mbox{sign}(k)\epsilon_{\mu\beta\gamma}p_\gamma-3\frac{p\cdot
n^*}{n\cdot n^*} \mbox{sign}(k)\epsilon_{\mu\beta\gamma}n_{\gamma}
\right.\nonumber\\
&&\left. +3\frac{p\cdot n^*}{(n\cdot n^*)(p\cdot
n)}\mbox{sign}(k)\left(n_\mu\epsilon_{\beta\gamma\delta}-n_\beta
\epsilon_{\mu\gamma\delta}\right)n_\gamma p_\delta +2 m\frac{n_\mu
n_\beta^*}{n\cdot n^*} \right],
\end{eqnarray}
so that
\begin{eqnarray}
\Omega_{{\rm (b)}\mu\nu}^{(1)ab}(p)
=\frac{1}{2\pi}g^2c_V\delta^{ab} \left[ \left(n\cdot
p\delta_{\mu\nu} -n_\mu p_\nu\right) -\frac{i}{2}m\,
\mbox{sign}(k)\epsilon_{\nu\beta\gamma}n_\gamma\frac{n_{\mu}
n_{\beta}^*}{n\cdot n^*}\right]. \label{ome2}
\end{eqnarray}
Adding Eqs.\,(\ref{valuea}) and (\ref{ome2}), we get
\begin{eqnarray}
\Omega_{\mu\nu}^{(1)ab}=\frac{1}{2\pi} g^2
c_V\delta^{ab}\left(n\cdot p\delta_{\mu\nu}-n_\mu p_\nu\right),
\label{omega}
\end{eqnarray}
where $\Omega_{\mu\nu}^{(1)ab}$, unlike $\Pi_{\mu\nu}(p)$ in
Eq.\,(\ref{vapo}), is obviously independent of $n_\mu^*$.
Inserting $\Omega_{\mu\nu}^{(1)}$  and $\Pi_{\mu\nu}(p)$ into the
relation (\ref{identity2}) and taking the limit $m\rightarrow
\infty$, we find that (\ref{identity2}) is indeed satisfied.

In this context, the following two points seem worth emphasizing.
First, looking at Eqs.\,(\ref{lre}) or (\ref{identity2}), we
notice again how inextricably the polarization tensor
$\Pi_{\mu\nu}$ is linked with the composite operator
$\Omega_{\mu\nu}^{(1)}$. Appearance of a non-zero
$\Omega_{\mu\nu}$ in the original identities (\ref{green}) and
(\ref{relation}) is clearly a signal of vector supersymmetry
breaking. This observation brings us to our second point, namely
the impact of the chosen regularization scheme  on the
corresponding symmetries. There are basically two possibilities:
we can (a) either adopt a regularization scheme which preserves
BRST invariance, but violates vector supersymmetry, or (b) employ
a scheme that preserves vector supersymmetry at the expense of
BRST symmetry.

The only meaningful choice, in our opinion, is to work with a
regularization procedure that respects BRST symmetry. The latter
unquestionably outranks vector supersymmetry, a topological
symmetry, in both overall effetiveness and field-theoretic
importance. Accordingly, the observed shift in the Chern-Simons
statistical parameter $k$ may be attributed  directly to a
BRST-preserving regularization scheme, in other words, to the
breaking of vector supersymmetry at the regularization level. A
similar conclusion holds for the anti-vector supersymmetry
(\ref{vector2}) and regardless whether the gauge is noncovariant
-- as in the present paper -- or covariant \cite{gmr,do}.

Having established the fact that there is no regularization in the
light-cone gauge that preserves both BRST and vector
supersymmetry, the question remaining is whether or not perhaps
the \underline{renormalized} effective action could be made both
BRST and vector supersymmetry invariant. To answer this question
we are guided  by the arguments given in \cite{gmr} for the Landau
gauge, where it was demonstrated that the quantum effective action
could indeed be expressed in an explicit BRST and vector
supersymmetry invariant form by an appropriate definition of the
fields. To appreciate this line of reasoning, we recall what
happens in the simple renormalization scheme for a finite theory,
defined by
\begin{eqnarray}
\Phi_R=Z_\Phi^{-1/2} \Phi, ~~~\Phi=(A_\mu^a, B^a, c^a,
\overline{c}^a),
\end{eqnarray}
with
\begin{eqnarray}
Z_A=Z_B^{-1}=Z_c=Z_{\overline{c}}= 1.
\end{eqnarray}
Here, the one-loop quantum effective action is given by
\cite{lema}
\begin{eqnarray}
\Gamma_R=\left(1+\frac{C_Vg^2}{4\pi}\right)S_{\rm CS}[A_R]+\int
d^3x\left(B^a_R n_\mu A_{R\mu}^a+\bar{c}^a_R n_\mu
D_{R\mu}^{ab}c_R^b\right), \label{qaction1}
\end{eqnarray}
where the subscript $R$ denotes the renormalized quantity,
$D_{R\mu}^{ab}=\partial_\mu\delta^{ab}+g f^{abc}A_{R\mu}^c$,
 and $Z_\Phi$ is the wave function renormalization
constant of the $\Phi$ field. In this particular renormalization
scheme, where the bare quantities equal the renormalized ones, the
quantum effective action (\ref{qaction1}) can only be BRST
invariant. However, if we choose the alternative renormalization
scheme defined by
\begin{eqnarray}
\Phi_{\widetilde{R}}=\widetilde{Z}_\Phi^{-1/2} \Phi,
\end{eqnarray}
with
\begin{eqnarray}
&&\widetilde{Z}_A=\widetilde{Z}_B^{-1}
=\widetilde{Z}_c=\widetilde{Z}_{\overline{c}}^{-1}
=1-\frac{g^2C_V}{4\pi},
\end{eqnarray}
the one-loop quantum effective action will become
\begin{eqnarray}
\Gamma_{\widetilde{R}}&=& S_{\rm CS}[A_{\widetilde{R}}] +\int
d^3x\left(B^a_{\widetilde{R}}n_\mu A_{\widetilde{R}\mu}^a+
\overline{c}_{\widetilde{R}}^an_\mu
D_\mu^{ab}c_{\widetilde{R}}^b\right)
\nonumber\\
&-&\frac{1}{8\pi}g^2C_V\int d^3x\left(\frac{1}{3!}\mbox{sign(k)}g
f^{abc}\epsilon_{\mu\nu\rho} A_{\widetilde{R}\mu}^a
A_{\widetilde{R}\nu}^bA_{\widetilde{R}\rho}^c -g
f^{abc}\overline{c}_{\widetilde{R}}^an_\mu A_{\widetilde{R}\mu}^b
c_{\widetilde{R}}^c\right). \label{qaction2}
\end{eqnarray}
It can be shown that the effective action $\Gamma_{\widetilde{R}}$
satisfies the Ward identities corresponding to both BRST symmetry
and vector (anti-vector) supersymmetry.

\vspace{2mm}

\begin{flushleft}
{\bf 5.~Conclusion}
\end{flushleft}

\vspace{2mm}

In this article, we have examined the origin of finite quantum
corrections in perturbative Chern-Simons theory in the
noncovariant light-cone gauge, $n\cdot A^a(x)=0$, $n_\mu
=(n_0,{\bf n})$, $n^2=0$. Our analysis consisted of five basic
steps:
\begin{itemize}
\item Variation of the Chern-Simons effective action $S$, namely
$v_\nu S$, where $v_\nu$ is the supersymmetric vector operator
defined in Eq.\,(\ref{vector1}).
\item Application of a gauge-invariant regularization scheme which
defines Chern-Simons theory as the large mass limit of
topologically massive Yang-Mills theory. The chosen regularization
scheme consists of dimensional regularization and the higher
covariant-derivative term $S_{\rm YM}$ in Eq.\,(\ref{ym}). This
hybrid regularization preserves BRST symmetry, but violates the
vector supersymmetry in Eq.\,(\ref{vector1}).
\item Use of the $n_\mu^*$-prescription in handling the spurious
singularities of $\left(n\cdot q\right)^{-1}$ in the gauge and
ghost propagators, where $n_\mu^*=\left(n_0, -{\bf n}\right)$.
\item Derivation of the Ward identity (\ref{green}) in coordinate
space or, equivalently, Eq.\,(\ref{relation}) in momentum space.
\item Discussion on the BRST and vector supersymmetry
invariance of the quantum effective action in the light-cone gauge
in the spirit of Ref.\,\cite{gmr}.
\end{itemize}
Our results may be summarized as follows:
\begin{enumerate}
\item The Ward identity in Eq.\,(\ref{relation}) was shown
to be satisfied both at the tree level, Eq.\,(\ref{tre}), and at
the one-loop level, Eq.\,(\ref{identity2}).
\item We have demonstrated that the composite operator
$\Omega_{\mu\nu}(p)$, which reflects vector symmetry breaking, is
inextricably linked with the vacuum polarization tensor
$\Pi_{\mu\nu}$ in Eq.\,(\ref{vapo}). We note that the latter is UV
finite and transverse, but contains one non-local, gauge-dependent
term.
\item The finite shift in the Chern-Simons statistical parameter
$k$, i.e. the finite quantum correction of the coupling constant,
was shown to arise specifically from the breaking of vector
supersymmetry.
\item The quantum effective action can be defined as being both
BRST invariant and light-cone vector supersymmetric invariant with
an appropriate choice of renormalization scheme.
\end{enumerate}

Conclusions similar to the above also hold for other
BRST-invariant regularization procedures, such as the hybrid
scheme consisting of higher covariant-derivative regularization
plus Pauli-Villars regularization \cite{do}. Nevertheless, it must
be emphasized that the light-cone gauge is particularly well
suited in pinpointing the origin of quantum corrections. The
reason can be found by looking at the two terms on the right-hand
side of the identity (\ref{relation}), both containing the ghost
propagator $S(p)$. Since $S(p)$ receives {\it no} quantum
corrections in the light-cone gauge, the first term proportional
to $\epsilon_{\mu\nu\rho}n_\rho S(p)$ maintains it classical
value. Consequently, the observed quantum corrections must
necessarily originate from the second term, specially from the
composite operator $\Omega_{\rho\nu}(p)$. No such conclusion is
possible in the Landau gauge, since quantum corrections are now
also generated in the ghost propagator $S(p)$.

 \section*{acknowledgment}
We should like to thank C.P. Martin for useful comments and
enlightening discussions. This work was supported in part by the
Natural Sciences and Engineering Research Council of Canada under
grant No.\,A8063.

\begin{figure}
\centering
\input FEYNMAN

\begin{picture}(40000,10000)
\THICKLINES \drawvertex\gluon[\S 3](5000,8000)[2]
\put(5200,\vertexoney){${\mu},a$}
\put(8000,\vertextwoy){${\rho},c$}
\put(0,\vertexthreey){${\nu},b$} \put(5000,200){(a)}

\drawvertex\gluon[\S 4](16000,8000)[3]
\put(\vertexonex,\vertexoney){${\mu},a$}
\put(\vertextwox,\vertextwoy){${\rho},d$}
\put(\vertexthreex,\vertexthreey){${\lambda},c$}
\put(11200,\vertexfoury){${\nu},b$} \put(16000,200){(b)}

\drawline\gluon[\S\REG](30000,8000)[2]
\put(\gluonfrontx,\gluonfronty){${\mu},a$}
\drawline\scalar[\SW\REG](\particlebackx,\particlebacky)[3]
\drawarrow[\SW\ATBASE](\pmidx,\pmidy)
\put(25000,\scalarbacky){$b$}
\drawline\scalar[\SE\REG](\gluonbackx,\gluonbacky)[3]
\drawarrow[\SE\ATBASE](\pmidx,\pmidy)
\put(34500,\scalarbacky){$c$} \put(30000,200){(c)}
\end{picture}
\caption{\protect\small Vertices of topologically massive
Yang-Mills theory: the wavy line represents the gauge field and
the dashed line represents the ghost field.}

\begin{picture}(40000,8000)
\THICKLINES \drawline\gluon[\SW\REG](5000,6000)[3]
\drawline\scalar[\SE\REG](\gluonfrontx,\gluonfronty)[3]
\put(4700,5900){$\bigotimes$}
\put(4800,50){$\widetilde{V}_{\nu\mu}^{ab}$}

\THICKLINES
 \drawline\scalar[\S\REG](16000,6000)[2]
\drawline\gluon[\SW\REG](16000,6000)[3]
\drawline\gluon[\SE\REG](16000,6000)[3]
\put(15800,5900){$\bigotimes$}
\put(16000,50){$\widetilde{V}_{\nu\mu\rho}^{abc}$}

\THICKLINES
 \drawline\gluon[\W\REG](30000,6000)[4]
\drawline\gluon[\SW\REG](\gluonfrontx,\gluonfronty)[3]
\drawline\gluon[\SE\REG](\gluonfrontx,\gluonfronty)[3]
\drawline\scalar[\E\REG](\gluonfrontx,\gluonfronty)[3]
\put(29900,5900){{\bf $\bigotimes$}}
\put(29900,50){$\widetilde{V}_{\nu\mu\rho\lambda}^{abcd}$}
\end{picture}

\caption{\protect\small Composite vertices relevant to vector
supersymmetry-breaking terms }

\begin{picture}(10000,5000)
\THICKLINES \drawline\scalar[\E\REG](0,0)[2] \drawloop\gluon[\NE
3](\pbackx,\pbacky) \drawline\scalar[\E\REG](\pbackx,\pbacky)[2]
\drawline\scalar[\W\REG](\pbackx,\pbacky)[4]
\end{picture}
\caption{\protect\small Self-energy of ghost field }

\begin{picture}(50000,22000)
\THICKLINES \drawloop\gluon[\NE 8](5000,18000)
\drawline\gluon[\W\FLIPPEDCENTRAL](6200,15100)[4]
\put(6300,15200){$\bigotimes$}
\drawline\scalar[\E\REG](6400,15200)[3] \put(6500,13050){(a)}

\drawloop\gluon[\NE 8](20000,17000)
\put(24400,16300){$\bigotimes$}
\drawline\gluon[\W\REG](20000,16300)[4]
\drawline\scalar[\E\REG](25200,16400)[2] \put(21000,13050){(b)}

\THICKLINES \put(35000,16000){$\bigotimes$}
\drawline\gluon[\W\REG](35000,16000)[3] \drawloop\gluon[\NE
3](\pfrontx,\pfronty) \drawline\scalar[\E\REG](35500,16000)[4]
\put(36000,13050){(c)}

\THICKLINES \drawline\gluon[\SW\REG](9000,10000)[3]
\put(8600,10100){$\bigotimes$}
\drawline\scalar[\SE\REG](\gluonfrontx,\gluonfronty)[3]
\drawline\gluon[\W\REG](\scalarbackx,\scalarbacky)[10]
\drawline\scalar[\E\REG](\gluonfrontx,\gluonfronty)[2]
\put(8600,1000){(d)}

\drawline\gluon[\SW\REG](32000,10000)[4]
\put(31600,10100){$\bigotimes$}
\drawline\scalar[\SE\REG](\gluonfrontx,\gluonfronty)[3]
\drawline\scalar[\W\REG](\scalarbackx,\scalarbacky)[6]
\drawline\gluon[\E\REG](\scalarfrontx,\scalarfronty)[3]
\put(31600,1000){(e)}
\end{picture}
\caption{\protect\small Feynman diagrams contributing to
$\Omega_{\mu\nu}^{(1)}$. Diagrams (c), (d) and (e) vanish. }

\end{figure}

\end{document}